# The effect of the magnetically dead layer on the magnetization and the magnetic anisotropy of the dextran-coated magnetite nanoparticles

*Zhila Shaterabadi*[1,2] · *Gholamreza Nabiyouni*[1,2] · *Gerardo F Goya*[3] · *Meysam Soleymani*[4]

1. Department of Physics, Faculty of Science, Arak University, Arak, 38156-88349, Iran
2. Department of Chemical Engineering, Faculty of Engineering, Arak University, Arak, 38156-88349, Iran
3. Condensed Matter Physics Department, Faculty of Sciences & Institute of Nanoscience and Materials of Aragón, University of Zaragoza, Zaragoza, Spain.
4. Institute of Nanoscience and Nanotechnology, Arak university, Arak, Iran

**Abstract**

We present a study on the magnetic behavior of dextran-coated $Fe_3O_4$ magnetic nanoparticles (Dex-M NPs) with sizes between 3 and 19 nm, synthesized by hydrothermal-assisted co-precipitation method. The decrease of saturation magnetization $M_s$ with decreasing particle size has been modeled by assuming the existence of a spin-disordered layer at the particle surface, which is magnetically dead. Based on this core-shell model, the dead layer thickness (t) and saturation magnetization ($M_S$) of the magnetic cores in our samples were estimated to be t = 6.8 Å and $M_s$ = 98.8 emu/g, respectively. The data of $M_s$ was analysed using a law of approach to saturation, indicating an increase in effective magnetic anisotropy ($K_{eff}$) with decreasing the particle size as expected from the increased surface/volume ratio in small MNPs. The obtained $K_{eff}$ were successfully modeled by including an extra contribution from dipolar interactions due to the formation of chain-like structure of MNPs. The surface magnetic anisotropy ($K_s$) was estimated to be about $K_s = 1.04 \times 10^5$ J/m$^3$. Our method provides a simple and accurate way to obtain the $M_s$ core values in surface-disordered MNPs, a relevant parameter required for magnetic modeling in many applications.

**Keywords:** Magnetic dead layer, Core-shell model, Spin disorder, Surface magnetic anisotropy, Dextran-coated magnetite NPs.

1. Introduction

Particle size reduction to the nanometer scale significantly affects magnetic properties of magnetic nanoparticles (MNPs) due to deleterious impact of surface atoms on the effective magnetization of the MNPs [1, 2]. However, when optimal performance is seek in bio-applications such as magnetic hyperthermia therapy [3, 4], drug delivery [5, 6], and magnetic resonance imaging [7, 8], it is important to retain the magnetization values M at room temperature as close as possible to the corresponding bulk ones [9, 10]. Hence, it is important to address size-dependent changes of magnetic properties in nanometer-scaled particles.

As the size of the MNPs decreases below the micrometer-size range several new phenomena appear, including superparamagnetism [11, 12], reduced saturation magnetization [13, 14], non-saturated and open hysteresis loop at high magnetic fields [15, 16]. For a given material with effective magnetic anisotropy $K_{eff}$, the superparamagnetic behavior appears when the particle volume V is small enough that the thermal energy can overcome the anisotropy energy barrier $E = K_{eff} V$ separating magnetization easy axes [17, 18]. On the other hand, the reduced and non-saturating magnetization phenomena originate from spin-disordered configuration at the MNP surface, that can be explained by the core-shell model.

This core-shell configuration on MNPs consists of a spin-disordered shell, known as magnetic dead layer (due to its zero net magnetization), surrounding a core with ferro/ferri magnetic-ordered spins [19, 20]. Deterioration of magnetic order in the dead layer is originated from the surface effects in this region. In fact, structural distortions at the MNPs boundaries result in breaking atomic bonds and consequently frustrating exchange interactions between surface and core spins, which in turn lead to the orientation deviation of surface spins with respect to the core ones [19, 21-24]. In ferrites, like magnetite, exchange interactions occur through intermediation of oxygen ions (called super-exchange interactions), and therefore the presence of defects and impurities in surface sites or missing of oxygen ions can spread the spin-disordered region into the core [21-23].

Consequently, the decrease in $M_s$ with the size reduction, which is one consequence of surface effects, has been well described by considering a model in which the MNPs are composed with a core having bulk-like magnetic properties and a surrounding shell composed of a magnetically disordered layer. In fact, as the particle size decreases the impaired magnetic order of the surface layer increasingly determines the magnetic properties of a given MNP. For iron oxide NPs with 0.9 nm magnetic dead layer thickness, Kim et al. reported that 61.4% of spins in 12 nm-sized MNPs are magnetically disordered, while the figure increasingly reaches 99.4% in 2.2 nm-sized MNPs [25]. In addition, the decreasing trend of $M_s$ with the size reduction has been reported in many articles [26-33]. Nevertheless, to the best of our knowledge, only one experimentally-estimated value has ever been reported for the magnetic dead layer thickness of magnetite NPs, not considering the coating ligand-related effect on the measured magnetization [34].

In this work, variable-sized Dex-M-NPs (from 3.1 to 18.9 nm) synthesized for magnetic hyperthermia [35] were used to investigate the effect of size reduction on the magnetization behavior. Specifically, we estimated the magnetic dead layer thickness of Dex-M NPs using real values of $M_s$ at high magnetic fields for the magnetic part of Dex-M NPs by eliminating the weight contribution of non-magnetic coating layer to the whole magnetization. The obtained results can provide new insight into the modification of magnetic properties of MNPs, especially for applications in which an accurate determination of $M_s$ is required, for instance magnetic heating models, magnetic actuation, etc.

## 2. Results and discussion

The Dex-M NPs in the size range 3.1–18.9 nm were synthesized by combination of co-precipitation and hydrothermal methods. The details of experimental procedure and sample characterization are given elsewhere [35]. The synthesis conditions of samples as well as some samples characteristics (e.g. $M_s$, TEM particle size (D), and the percentage of remanence mass in TG analyses ($m_r$)) are summarized in Table 1. We note that our largest applied field in VSM measurements could be insufficient for complete saturation of samples, and thus the $M_s^*$ values were estimated using extrapolation of magnetization (M) versus the inverse of magnetic field strength (1/H) curves. To this end, initial magnetization curves of Dex-M-NPs (shown in Fig. 1) were utilized to make M $vs.$ $\frac{1}{H}$ curves using their data near saturation. Both the $M_s^*$ and $M_s$ values are presented per unit of total mass ($g_{Dex-M}$) which comprises the mass of magnetite NPs ($g_{Mag}$), dextran layer, and absorbed water on NPs surface. Therefore, $M_s^{**}$ values (real saturation magnetizations at high magnetic filed strengths for pure magnetite NPs) were estimated using effective magnetic material mass obtained by $m_r$ in TG analyses. The values of $M_s^*$, and $M_s^{**}$ are also summarized in the Table 1.

Table 1  Synthesis conditions, TEM particle size, and some characteristics of the Dex-M NPs

| Sample | Synthesis conditions* | D (nm) | $M_s$ (emu/$g_{Dex-M}$) | $m_r$ (%) | $M_s^*$ (emu/$g_{Dex-M}$) | $M_s^{**}$ (emu/$g_{Mag}$) |
|---|---|---|---|---|---|---|
| Dex-M-80  | Co 80°C         | 3.1±0.4  | 8.3  | 60.73 | 9.9  | 16.3 |
| Dex-M-120 | Co80°C+Hy120°C  | 4.5±0.4  | 26.5 | 72.66 | 27.9 | 38.4 |
| Dex-M-140 | Co80°C+Hy140°C  | 6.7±0.3  | 33.8 | 80.60 | 34.9 | 43.3 |
| Dex-M-160 | Co80°C+Hy160°C  | 8.1±0.2  | 44.4 | 82.21 | 45.3 | 55.1 |
| Dex-M-180 | Co80°C+Hy180°C  | 11.5±0.2 | 59.2 | 82.62 | 59.9 | 72.5 |
| Dex-M-200 | Co80°C+Hy200°C  | 15.0±0.3 | 64   | 86.38 | 64.7 | 74.9 |
| Dex-M-220 | Co80°C+Hy220°C  | 18.9±0.3 | 67.9 | 87.69 | 68.4 | 78   |

* Co-precipitation and hydrothermal synthesis methods are respectively written as Co and Hy for short. Also, the synthesis temperature is written next to each method name.

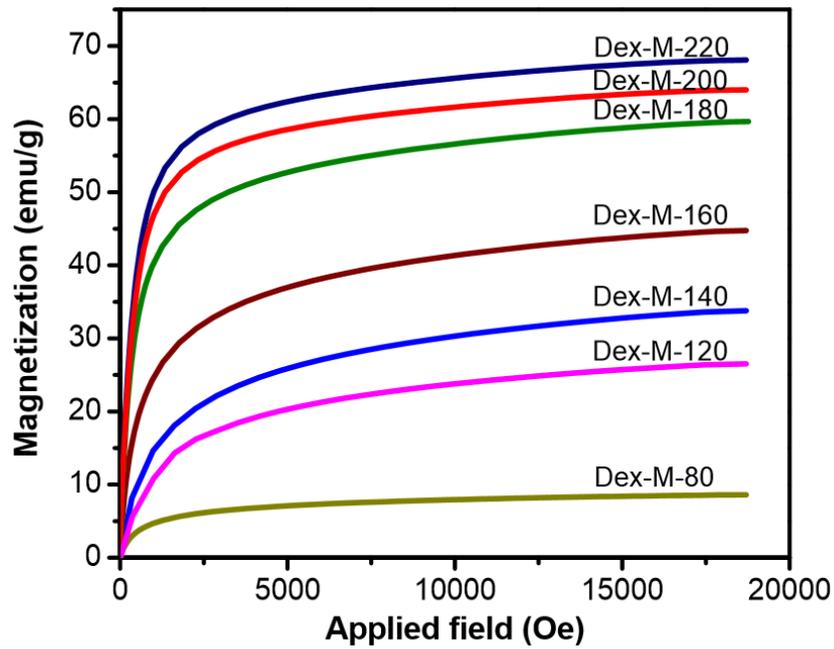

Fig. 1 Initial magnetization curves of Dex-M-NPs as a function of applied magnetic field

As it can be seen in the Table 1, $M_s^{**}$ values are smaller than those of the bulk material in the range $84 - 100$ emu/g [17, 33, 36-38]. Moreover, they dramatically decrease with the particle size reduction. Decline in the saturation magnetization with decrease in the particle size, which has already been observed in other experimental works [39-43], can be originated both from redistribution of cations between two sub-lattices of spinel structure and spin disorder on the particle surface. It has been reported that the distribution of cations between tetrahedral and octahedral sites of spinel structure can significantly affect the magnetization of ferrite NPs [76]. Magnetite ($Fe_3O_4$) is an interesting member of the spinel ferrite family with inverse structure as $(Fe^{3+})(Fe^{2+}Fe^{3+})O_4$ in which the parentheses respectively indicate the tetrahedral and octahedral sites [4,76]. Consideration the magnetic moments of $Fe^{3+}$ and $Fe^{2+}$ ions as 5 and 4 $\mu_B$ respectively, the net magnetic moment of each magnetite molecule is simply calculated as 4 $\mu_B$. Assuming the probable redistribution of a fraction (x) of cations, a partially deviated inverse spinel structure as $(Fe^{2+}_xFe^{3+}_{1-x})(Fe^{2+}_{1-x}Fe^{3+}_{1+x})O_4$ with the net magnetic moment $(4 + 2x)$ $\mu_B$ is formed [44].

Accordingly, even if the size reduction causes a change in the arrangement of cations in the spinel structure of the magnetite NPs, magnetization is expected to increase. In other words, the magnetization reduction can exclusively be attributed to the existence of the magnetically inert layer on the surface of the MNPs in the core-shell model.

Assuming that the magnetic dead layer has a) a negligible net magnetization and b) a thickness, t, independent of particle size, D, the saturation magnetization $M_s$ is given by

$$M_s = M_{s0} (1 - 2t/D)^3 \qquad (1)$$

where $M_{s0}$ is the saturation magnetization of bulk material. Ec.(1) indicates that the decrease of our experimental values of $M_s^{**}$ (i.e., the magnetization corrected for the dextran mass) should be most relevant for the smallest MNPs. The plot of $M_s^{**1/3}$ vs. $\frac{1}{D}$ data (Fig. 2) could be well fitted by a linear function as expected from Ec. (1), obtaining the values of $M_{s0} = 98.8$ emu/g and t = 6.8 Å for the bulk saturation magnetization and dead layer thickness, respectively ($R^2 = 0.9662$).

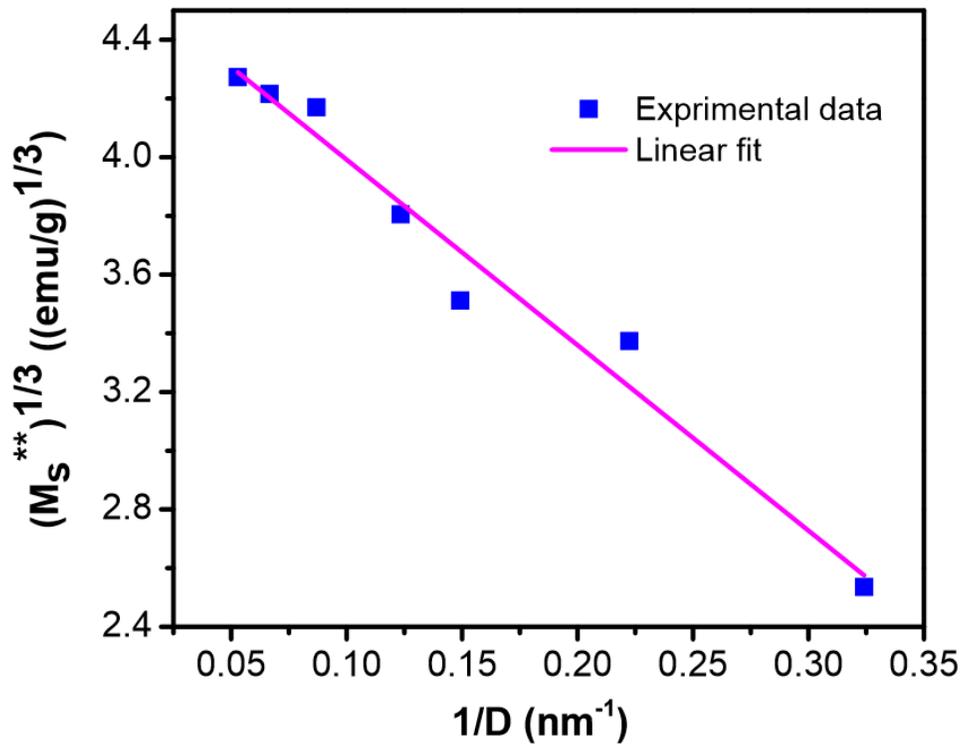

Fig. 2 The cube root of saturation magnetization ($M_s^{**1/3}$) versus the inverse of average diameter (1/D) of Dex-M-NPs

The $M_{s0}$ obtained from the fit is consistent with reported values of bulk magnetite (95.0-98.5 emu/g) [17, 33, 36-38]. The value of the thickness t = 6.8 Å obtained from the fit is somewhat smaller than the $Fe_3O_4$ lattice constant (8.39 Å) [45] but comparable to previous findings $t = 6$ Å found by Chen et al. [46], and Zheng et al. [47] for $MnFe_2O_4$ NPs at 300 K. Other $t$ values consistent with our findings have been reported in different systems, including values of $t = 5$Å (at 10 K) [46], and 4.5 Å (at 20 K) [48] for $MnFe_2O_4$ MNPs, and 10 Å (at 5 K) for $CoFe_2O_4$ NPs [49]. Interestingly, $CuFe_2O_4$ NPs with sizes from 10 to 60 nm produced by mechanical milling were reported to have a much larger t ≈ 2.5 nm spin-disordered layer thickness, consistent with the high-energy collisions during mechanosynthesis [50]. The only measured values for the thickness $t$ in $Fe_3O_4$ MNPs, to the best of our knowledge, has been reported by Caruntu et al. to be $t = 2.26$ Å (at 300 K) and $t = 1.92$ Å (at 5 K) [34]. Although these values are smaller than the estimation presented here, comparison between samples obtained from different synthesis routes can be misleading, since the thickness of the disordered surface is expected to depend on the details of the energy landscape involved in each synthesis route. We also note that the thickness t obtained from the extrapolation method could be affected by the $M_s$ values assumed for the magnetic core at high magnetic filed. Our estimations are based on considering the $M_s^{**}$ value obtained from the extrapolation of M vs.1/H curves and the TG results.

Using the obtained t value and assuming the spherical shape for the Dex-M NPs, the magnetic-disordered content for the Dex-M-80 NPs was calculated to be 82.3 %. It means that the magnetic order only exists in 17.7% of the particle volume. By increasing the particle size, the magnetic-ordered content increases to 79.9 % for the Dex-M-220 NPs. Therefore, the effect of surface magnetic disorder on the magnetic behavior of MNPs significantly diminishes with increasing the particle size, consistent with the decrease of the surface/volume ratio as the particle size increases.

The magnetization data as a function of the applied field H can be analyzed in terms of the law of approach to saturation [51]:

$$M = M_S \left(1 - \frac{b}{H^2}\right) \qquad (2)$$

where the parameter b is related to the magnetocrystalline anisotropy, whichcan be obtained from the M(H) data near saturation by a linear fitting of the $\frac{M}{M_s}$ (or $\frac{M^{**}}{M_S^{**}}$ in our assumption) versus $1/H^2$ curve. Following the procedure reported in [52], we calculated $K_{eff}$ for a uniaxial magnetic anisotropy as Eq. (3) [53, 54].

$$K_{eff} = \mu_0 M_S \left(\frac{15}{4}b\right)^{1/2} \qquad (3)$$

Using the b values obtained from Eq. (2) and the $M_S^{**}$ values from Table 1, Eq. (3) yielded the $K_{eff}$ values for Dex-M NPs listed in Table 2. These calculated values are higher than the bulk anisotropy constant ($K_{bulk\ magnetite} = 1.35 \times 10^4$ J/m$^3$ [55]), reflecting the surface effects as particle size decreases. This size dependence is consistent with previous findings in iron oxide NPs [34, 52, 55, 56] that reported a rising trend for $K_{eff}$ with size reduction. The size dependency of the $K_{eff}$ is in good agreement with previous results on spherical Fe$_3$O$_4$ MNPs showing a decrease in $K_{eff}$ from $4.74 \times 10^5$ J/m$^3$ to $1.11 \times 10^5$ J/m$^3$ with increasing particle size from 6 to 11 nm [34]. Similar changes in the $K_{eff}$ for cubic magnetite NPs have been observed, with a reduction from $77 \times 10^3$ J/m$^3$ (20 nm diameter) to $42 \times 10^3$ J/m$^3$ (40 nm diameter) [52]. Sarkar and Mandal also reported a decreasing trend in the $K_{eff}$ from $1.84 \times 10^5$ J/m$^3$ (7.23 nm diameter) to $1.25 \times 10^5$ J/m$^3$ (11 nm diameter) for chain-like magnetite NPs [55].

We mention that, since the overall shape of our MNPs does not change significantly along the series of samples as observed from TEM images,[35] the contributions from shape anisotropy to $K_{eff}$ can be ignored.

Assuming that the magnetocrystalline anisotropy $K_{bulk}$ of the magnetic cores is constant along our series of samples with different particle sizes, additional contributions to the *effective* magnetic anisotropy $K_{eff}$ come from shape and/or surface as well as magnetic dipolar interactions among MNPs [57-59]. Our experimental determination of the evolution in both $K_{eff}$ and $K_S$ was made on non-diluted samples, so dipolar magnetic interparticle interactions could be not negligible in the analysis of the single-particle magnetic anisotropy. However, the dextran-coating makes the MNPs to be separated by (at least) a distance twice the coating layer thickness, and therefore dipole-dipole interactions between particles can be assumed to be constant along our sample series [34, 60, 61]. Additionally, since the $M_s$ values in our MNPs decrease with decreasing particle size (see Table 1) the same trend are to be expected for the strength of dipole-dipole interactions [61].

The phenomenological expression for the anisotropy $K_{eff}$ originally proposed by Bødker et al. [62],

$$K_{eff}V \cong K_{bulk}V + K_s S \qquad (4)$$

where $K_{bulk}$ is the bulk anisotropy energy per unit volume, and $K_s$ is the surface density of anisotropy energy. Assuming that the particles are spherical with diameter D, Bødker et al. simplified the Eq. (4) as Eq. (5) which has been experimentally found on many different systems [63, 64].

$$K_{eff} \cong K_{bulk} + \frac{6K_s}{<D>} \qquad (5)$$

Using symmetry arguments and assuming that surface anisotropy is normal to the particle surface, Bødker et al. [62] showed that for a perfectly spherical particle a zero contribution from surface anisotropy should be expected. We note here that this is an empirical expression, and the hypothesis that the surface contribution to the effective anisotropy is simply additive has yet to be demonstrated.

Fig. 3(a) shows fitting the $K_{eff}$ vs. 1/D data using Eq. (5), from which a value $K_s = 2.11 \times 10^5$ J/m$^3$ was obtained for our Dex-M NPs samples. However, as clearly seen in Fig. 3(a), there is a large deviation from the linear behaviour for large 1/D values. These deviations could be explained by deviations from spherical shape that are not included in Eq. (5). Indeed, for different particle morphologies Eq. (4) should include an additional contribution with a different $K_{eff}$ $vs. \frac{1}{D}$ slope. However, as previously mentioned no major change in MNPs morphology can be observed in our series of increasing-size samples, in spite of the hydrothermal route used.[35, 65-67].

It should be noted that in the original work by Bødker et al., did not consider any contributions from dipolar interactions to the collective behavior of nanoparticles. It is well known that dipolar interactions between MNPs favor the formation of the chain-like structure in large enough MNPs for which the energy of the magnetic dipole-dipole interaction energy can surpass the thermal energy even at room temperature [68-73]. To include these interactions, we modified Eqs. 4 and 5 including a parabolic term

$$K_{eff} \cong K_{bulk} + \frac{6K_s}{<D>} + \frac{E_l}{<D^2>} \qquad (6)$$

in which $E_l$ has units of a linear density of anisotropy energy. Fig. 3(b) clearly shows that using Eq. (6) the fit of experimental data can be extended to the full range of particle sizes. The last term in Ec. (6) could be understood as originated in the formation of one-dimensional chains of MNPs (i.e. head-to-tail orientation) due to dipolar interactions. Consistent with Ec.(6) the last term is more relevant for larger 1/D values, reflecting the fact that chain formation is favoured for larger MNPs due to their larger dipolar moment. This interaction originates the extra contribution to the anisotropy in the Eq. (6) and, with this assumption and using Eq. (6) to fit the $K_{eff}$ vs. 1/D data, a value of $K_s = 1.04 \times 10^5$ J/m³ was obtained for our Dex-M NPs samples. The obtained value shows that surface anisotropy gives an important contribution to the effective anisotropy of small MNPs. In fact, surface anisotropy originates from the lack of long-range crystalline order in surface layer where breaking the crystal structure symmetry due to the lower and more variable coordination of cations results in perturbation in crystal field and consequently modification of magnetocrystalline anisotropy. [21, 23].

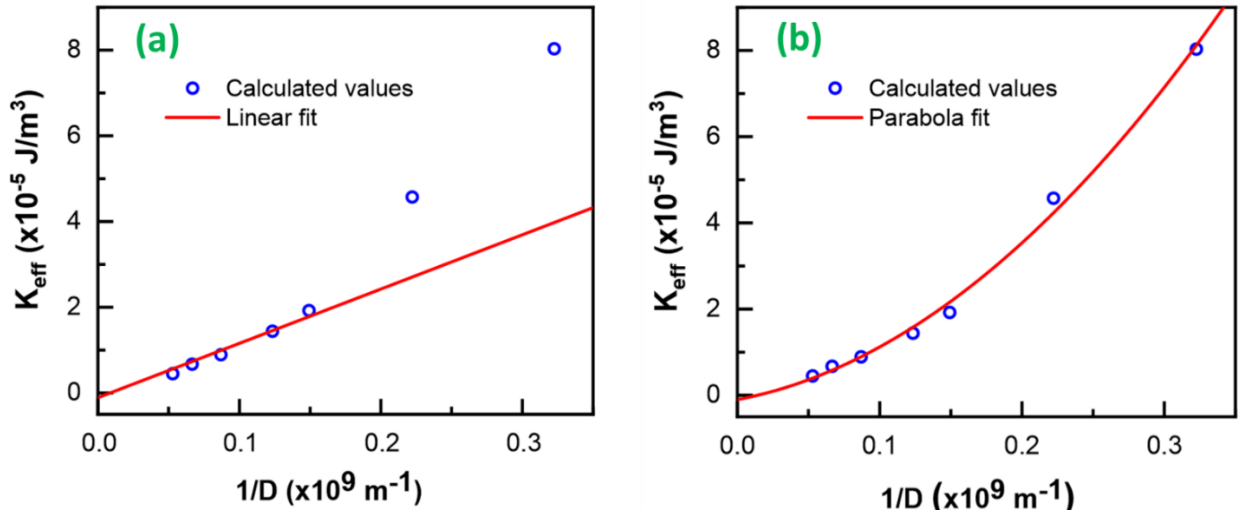

Fig. 3 fitting the $K_{eff}$ vs. 1/D data using (a) Eq. (5) and (b) Eq. (6)

Table 2  The calculated $K_{eff}$ values of the Dex-M-XX nanoparticles.

| Sample Dex-M-XX | M-80 | M-120 | M-140 | M-160 | M-180 | M-220 | M-220 |
|---|---|---|---|---|---|---|---|
| $K_{eff}$ (× 10⁵ J/m³) | 8.03 | 4.57 | 1.92 | 1.44 | 0.89 | 0.97 | 0.45 |

## 3. Conclusion

We have successfully used a series of magnetite ($Fe_3O_4$) nanoparticles of increasing sizes from 3.1 to 18.9 nm to investigate size-dependent changes in their magnetic properties. Our results revealed that the decrease in the saturation magnetization $M_s$ with decreasing size can be explained by a magnetically-disordered surface layer, and fitting the experimental data the values of the magnetic dead layer thickness and $M_s$ were estimated as $t = 6.8$ Å and $M_s = 98.8$ emu/g, respectively. We used a modified relation for calculating the contribution of the surface anisotropy $K_S$ to the effective anisotropy $K_{eff}$ by adding the contributions from dipolar interactions to the original model proposed by Bødker et al., obtaining a good fit for the whole range of MNPs sizes. Our analysis provides a more clear picture of the effects from the spin-disordered surface configuration on the magnetic properties in MNPs of diverse sizes. .